# Exploring van der Waals materials with high anisotropy: geometrical and optical approaches

*Aleksandr S. Slavich[1,†], Georgy A. Ermolaev[2,†], Mikhail K. Tatmyshevskiy[1], Adilet N. Toksumakov[1], Olga G. Matveeva[1], Dmitriy V. Grudinin[2], Arslan Mazitov[4], Konstantin V. Kravtsov[1], Alexander V. Syuy[2], Dmitry M. Tsymbarenko[3], Mikhail S. Mironov[2], Sergey M. Novikov[1], Ivan Kruglov[2], Davit A. Ghazaryan[1,5], Andrey A. Vyshnevyy[2], Aleksey V. Arsenin[2,5], Valentyn S. Volkov[2,5], and Kostya S. Novoselov[6,7,8,\*]*

[1]*Moscow Center for Advanced Studies, Kulakova str. 20, Moscow, 123592, Russia*

[2]*Emerging Technologies Research Center, XPANCEO, Dubai Investment Park First, Dubai, United Arab Emirates*

[3]*Department of Chemistry, Lomonosov Moscow State University, Moscow, 119991, Russia*

[4]*Institute of Materials, École Polytechnique Fédérale de Lausanne, 1015 Lausanne, Switzerland*

[5]*Laboratory of Advanced Functional Materials, Yerevan State University, Yerevan 0025, Armenia*

[6]*National Graphene Institute (NGI), University of Manchester, Manchester, M13 9PL, UK*

[7]*Department of Materials Science and Engineering, National University of Singapore, Singapore, 03-09 EA, Singapore*

[8]*Chongqing 2D Materials Institute, 400714, Chongqing, China*

[†]These authors contributed equally to this work

*Correspondence should be addressed to e-mail: kostya@nus.edu.sg

## Abstract

**The emergence of van der Waals (vdW) materials resulted in the discovery of their giant optical, mechanical, and electronic anisotropic properties, immediately enabling countless novel phenomena and applications. Such success inspired an intensive search for the highest possible anisotropic properties among vdW materials. Furthermore, the identification of the most promising among the huge family of vdW materials is a challenging quest requiring innovative approaches. Here, we suggest an easy-to-use method for such a survey based on the crystallographic geometrical perspective of vdW materials followed by their optical characterization. Using our approach, we found $As_2S_3$ as a highly anisotropic vdW material. It demonstrates rare giant in-plane optical anisotropy, high refractive index and transparency in the visible range, overcoming the century-long record set by rutile. Given these benefits, $As_2S_3$ opens a pathway towards next-generation nanophotonics as demonstrated by an ultrathin true zero-order quarter-waveplate that combines classical and the Fabry-Perot optical phase accumulations. Hence, our approach provides an effective and easy-to-use method to find vdW materials with the utmost anisotropic properties.**

**Main**

Modern nanophotonics exploits a plethora of novel phenomena for advanced light manipulation. Among them, are bound states in the continuum[1], chirality-preserving reflection[2], virtual-reality imaging[3,4], and others. The key parameter in these effects is the refractive index $n$ since it governs the resonance wavelength $\lambda_{\text{res}}$ ($\lambda_{\text{res}} \sim 1/n$)[5] and the resonance quality factor $Q$ (*e.g.*, $Q \sim n^2$ for the Mie resonances)[6] and the optical power, which is proportional to $n - 1$. Hence, even a slight increase in the refractive index gives a tremendous advantage in optical applications[7]. However, the refractive index is fundamentally limited[7], with the best results provided by high-refractive index materials, such as Si[8], GaP[9], TiO$_2$[10], InGaS$_3$[11], and SnS$_2$[12]. As a result, a new strategy of using anisotropic materials[13–19] has appeared in recent years due to the emergence of an additional degree of freedom for light manipulation, namely, optical anisotropy. Optical anisotropy revolutionizes integrated nanophotonics[20] by enabling subdiffractional light guiding[21,22], polariton canalization[23,24], Dyakonov surface waves[25], and the high integration density of waveguides[26,27].

The most promising anisotropic materials are van der Waals (vdW) crystals[15,28–30]. They are bulk counterparts of two-dimensional (2D) materials. Their 2D layered origin naturally leads to record values of optical anisotropy[15] because of the fundamental difference between in-plane covalent and out-of-plane vdW atomic bonds. However, it mostly results in out-of-plane birefringence, while some interesting effects require in-plane anisotropy[16,31]. At present, rutile continues to hold the record for strongest in-plane optical anisotropy in the visible range despite the significant advances in materials science[18,32], which is surprising considering that several decades have passed since its discovery[33]. It has inspired intensive research of low-symmetry vdW crystals[34–37].

In this work, we provide a solution to this long-term challenge. Our consideration of lattice geometry reveals that arsenic trisulfide (As$_2$S$_3$) stands out among other low-symmetry crystals. Further study of this material by micro-transmittance and spectroscopic ellipsometry measurements in combination with quantum mechanical computations confirmed that It possesses a giant in-plane optical anisotropy in the visible range. They also show that As$_2$S$_3$ belongs to transparent high-refractive index materials. Thus, As$_2$S$_3$ offers a universal material platform for nanooptics that brings benefits of both giant optical anisotropy and high refractive index.

**Origins of giant optical anisotropy of van der Waals materials**

Recent investigations[29,38] of vdW materials' optical properties reveal that those constitute the next-generation high refractive index materials platform with about 80% larger polarizability compared to traditional photonic materials, such as Si, GaP, and TiO$_2$. Hence, the search for highly refractive materials in the visible range among vdW crystals is a natural next step. Nevertheless, it is a tedious task because there are more than 5000 vdW crystals[39], and a straightforward enumeration of options is unacceptably time-consuming. To reduce the search area and pick the most promising materials for nanophotonics, we particularly aim for high in-plane optical anisotropy. Still, the family of anisotropic vdW crystals is huge, which motivates us to identify features relevant to large optical anisotropy. This problem is challenging since optical anisotropy could result from numerous unrelated physical effects. Among them are preferential directions of excitons[15,18,40], atomic-scale modulations[17], quasi-one-dimensional structures[16,41], different natures of atomic bonding[15], aligned interaction of dipole excitations around specific atoms[42], phonon resonances[14,43], and many others. Obviously, one of the reasons for the strong optical anisotropy is the directional material resonances, such as excitons[15,18,40] and phonons[14,43]. However, it is difficult to identify directional resonances or types of atomic bonding since it requires costly quantum-mechanical simulations. Therefore we choose an alternative approach, and assume that excitations with very strong anisotropy can manifest themselves in or be caused by the geometrical anisotropy of an elementary crystal cell.

Next, we compare the crystal structure of the most representative in-plane anisotropic crystals in Figure 1a. Interestingly, two materials stand out, namely, $Sr_{9/8}TiS_3$ and $As_2S_3$. According to Figure 1a, they exhibit the largest "geometric anisotropy", that is, the ratio of in-plane lattice parameters. Indeed, a recent study[17] shows that $Sr_{9/8}TiS_3$ has the largest optical anisotropy ($\Delta n \sim 2.1$) with zero losses, which coincides with our predictions. However, the optical bandgap of $Sr_{9/8}TiS_3$ corresponds to the infrared spectral range. In contrast, the optical bandgap of $As_2S_3$ is within the visible range with $E_g \sim 2.7$ eV ($\lambda_g \sim 460$ nm)[44,45], and $As_2S_3$ crystal (Figure 1b) demonstrates a similar to $Sr_{9/8}TiS_3$ "geometric anisotropy". Consequently, we anticipate that $As_2S_3$ will offer a giant optical anisotropy and, like other vdW materials with strong optical anisotropy[38], a high refractive index in the visible range.

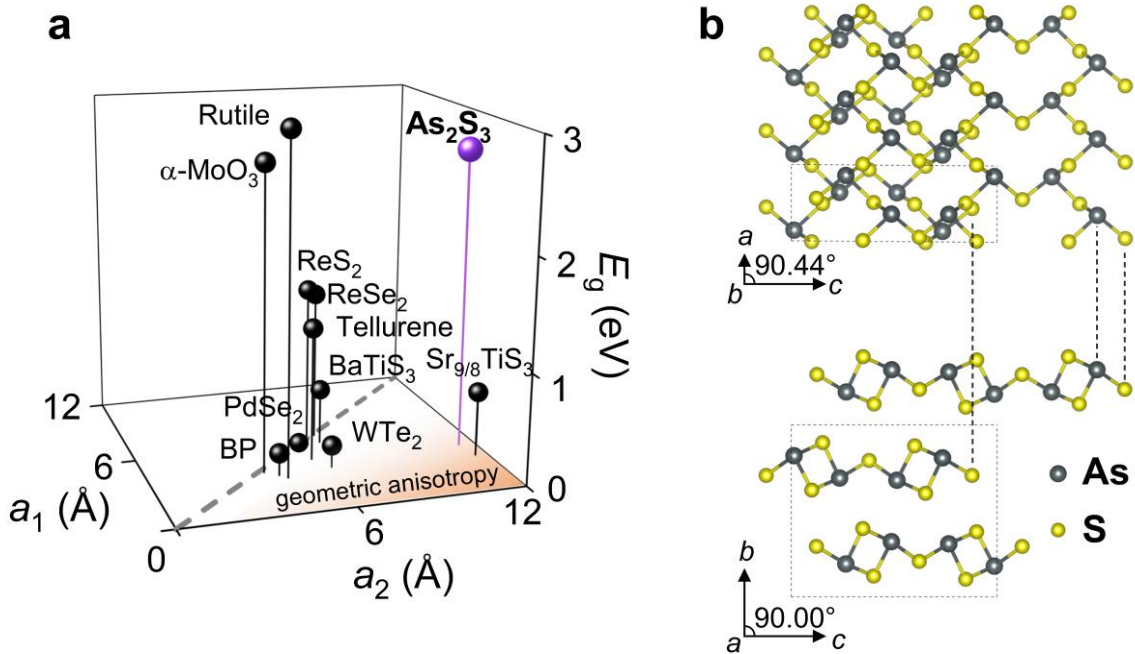

**Figure 1| Anisotropic crystalline structure as an origin for giant optical anisotropy. a,** The comparison of the lattice parameters of the representative anisotropic crystals and their bandgaps. $a_1$ and $a_2$ stand for the in-plane lattice parameters. **b,** Monoclinic crystal structure of $As_2S_3$ along the crystallographic b-axis (top) and a-axis (bottom). The black dashed frame shows the unit cell.

**Crystal structure of van der Waals $As_2S_3$**

$As_2S_3$ is a yellow semiconducting crystal usually found in nature as the mineral orpiment[44]. Amorphous $As_2S_3$ has already proved useful in such photonic applications as holography[46] and fibers[47]. At the same time, $As_2S_3$ in vdW configuration (see Figure 1b and Supplementary Note 1 for $As_2S_3$ characterization) appeared only recently in the research focus owing to the extraordinarily large in-plane mechanical anisotropy[48,49]. It also shows that our approach based on lattice geometry consideration in Figure 1a could be used to search for other giant anisotropic properties beyond optical ones.

In light of the importance of lattice parameters, we commenced our study of $As_2S_3$ with their refinement via X-ray diffraction measurements (see Methods). The XRD imaging patterns in Figures 2a-c confirm the monoclinic structure of $As_2S_3$ (see Figure 1b) with the following lattice parameters: $a = 4.2546(4)$ Å, $b = 9.5775(10)$ Å, $c = 11.4148(10)$ Å, $\alpha = 90°$, $\beta = 90.442(4)°$, and $\gamma = 90°$. Using these values of the parameters, we computed the bandstructure of $As_2S_3$ (see Supplementary Note 2) from the first principles (see Methods). Of great interest are the bandstructure cuts along crystallographic axes, presented in Figures 2d-f. First of all, we can notice a considerable difference in the dispersion curves, which clearly indicates significant anisotropic properties. Moreover, the bandstructure cuts for in-plane directions have a fundamental difference: along the *a*-axis, the bandstructure has a high dispersion, while it is flat along the *c*-axis. Therefore, we expect a stronger dielectric response for the *c*-axis than for the *a*-axis as flat bands lead to a large density of states and, as a result, a large refractive index[7,50].

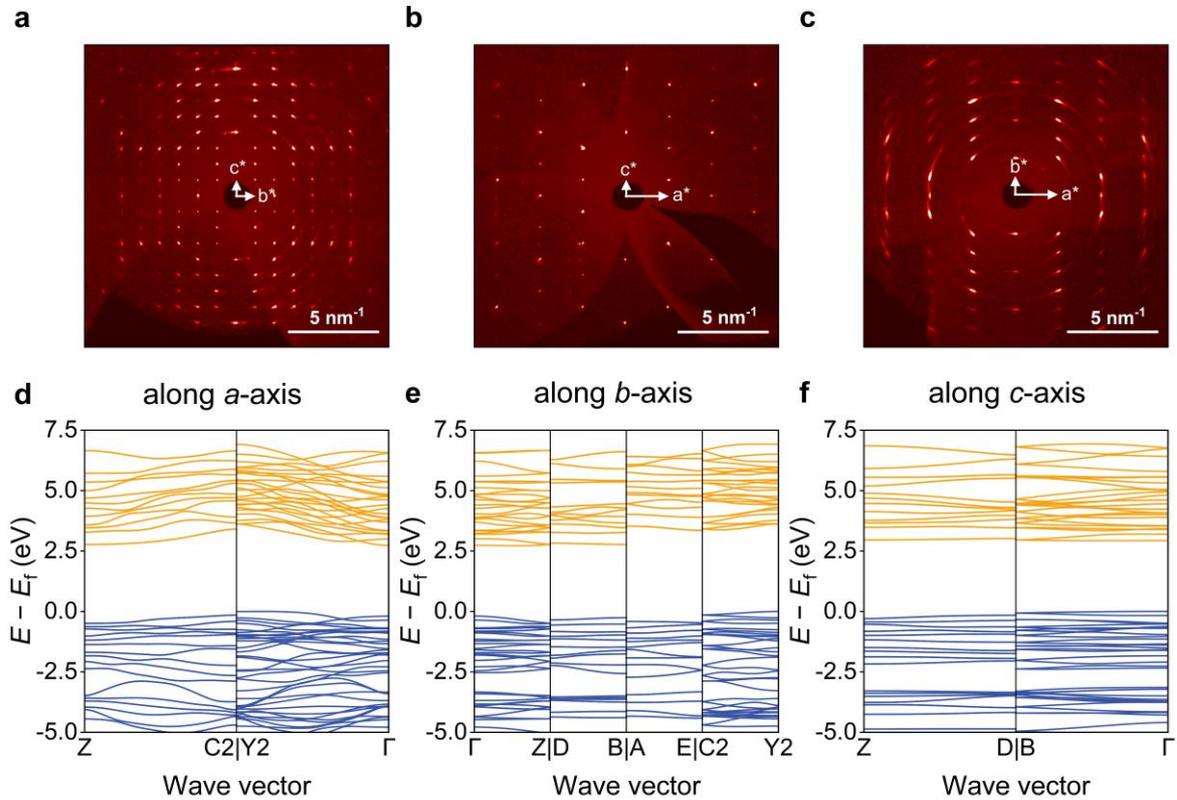

**Figure 2 | As$_2$S$_3$ anisotropic crystal structure, a comprehensive characterization.** XRD patterns in reciprocal space along **a,** *c\*-b\**, **b,** *c\*-a\**, and **c,** *b\*-a\** reciprocal planes. The electronic bandstructure cuts along **d,** *a*-axis, **e,** *b*-axis, and **f,** *c*-axis. Orange and blue curves present conduction and valence bands, respectively.

**In-plane optical anisotropy of van der Waals As$_2$S$_3$**

In general, for monoclinic systems, anisotropic permittivity tensor has a diagonal form diag($n_a$, $n_b$, $n_c$) in the crystallographic (*a*, *b*, *c*) basis, where $n_a$, $n_b$, and $n_c$ are refractive indices along the corresponding crystallographic axes[51]. The problem with this description is a non-orthogonal (*a*, *b*, *c*) basis, which significantly complicates the determination and the use of monoclinic optical constants due to the impossibility of decoupling the contribution of $n_a$, $n_b$, and $n_c$ into the optical response of the monoclinic crystal. Luckily, the monoclinic angle *β* of As$_2$S$_3$ differs from 90° very slightly by just 0.442(4)°, which allows us to treat As$_2$S$_3$ as an orthorhombic crystal. In this approximation, we can separately probe As$_2$S$_3$ optical components ($n_a$, $n_b$, and $n_c$) by orthogonal polarizations. For this purpose, we measured the polarized micro-transmittance of As$_2$S$_3$ flakes exfoliated on Schott glass substrates (see Methods) and determined their crystallographic axes by polarized Raman spectroscopy (see Supplementary Note 3). The exemplified transmittance spectra maps for parallel- and cross- polarizations for the case of 345-nm-thick flake are plotted in Figures 3a-b. Note that we choose the transparency range (500 – 850 nm) of As$_2$S$_3$, which allows us to leverage Cauchy models (see Methods) for As$_2$S$_3$ refractive indices[15]. Using this description, we fitted the experimental data (see Figures 3a-b), and calculated the corresponding spectra in Figures 3c-d. Calculations agree perfectly with the experiment (Figures 3a-b) and give us in-plane optical constants of As$_2$S$_3$ presented in Figure 3e. However, micro-transmittance cannot probe the out-of-plane component of the As$_2$S$_3$ dielectric tensor. Therefore, we performed single-wavelength Mueller-Matrix spectroscopic ellipsometry and near-field studies (see Supplementary Notes 4-6) presented in Figure 3e to get the complete picture of the As$_2$S$_3$ dielectric response. Additionally, we performed the first-principle calculations (see Methods) of the As$_2$S$_3$ dielectric function (see Figure 3e), which coincides well with the measured values, especially for in-plane components, $n_a$ and $n_b$. Notably, even first-principle computations yield zero extinction coefficient $k$ for the considered visible range (see Figure 3e), which confirms that As$_2$S$_3$ is a lossless material, promising for visible nanophotonics

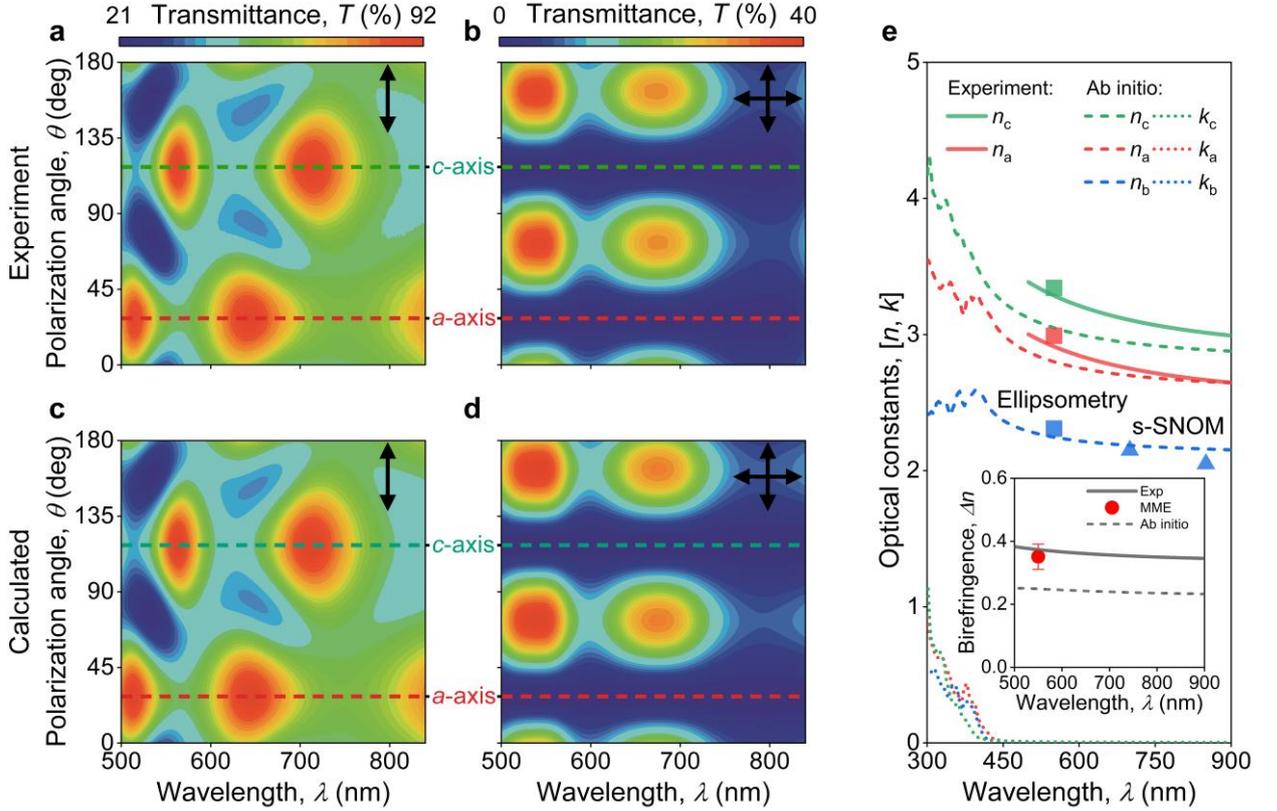

**Figure 3 | Optical properties of van der Waals As$_2$S$_3$.** Experimental polarized micro-transmittance of an As$_2$S$_3$ flake for **a,** parallel- and **b,** cross-polarized configurations. Calculated polarized micro-transmittance of As$_2$S$_3$ for **c,** parallel- and **d,** cross-polarized configurations. The dashed lines show the position of crystallographic axes $a$ (red line) and $c$ (green line). **e,** Anisotropic optical constants of As$_2$S$_3$. The inset shows the in-plane birefringence of As$_2$S$_3$. Tabulated optical constants of As$_2$S$_3$ are collected in Supplementary Note 7.

**As$_2$S$_3$ in the family of high-refractive index materials**

Of immediate interest are absolute values of As$_2$S$_3$ optical constants (see Figure 3e and Figures 4a-b). As anticipated from the bandstructure calculations in Figures 2d-f, As$_2$S$_3$ has the largest refractive index $n_c$ along the crystallographic $c$-axis (see Figure 3e). Besides, the benchmarking of $n_c$ with other crystals (Figure 4a) reveals that As$_2$S$_3$ also belongs to the family of high refractive index materials and holds record values below 620 nm. If we extrapolate the Cauchy model for $n_c$ to As$_2$S$_3$ optical bandgap ($E_g \sim 2.7$ eV), then we can clearly see that As$_2$S$_3$ fits the correlation for vdW materials between the optical bandgap and its refractive index, as shown in Figure 4c.

Apart from a high refractive index, As$_2$S$_3$ possesses giant in-plane optical anisotropy $\Delta n \sim 0.4$ (see the inset in Figure 3e). This is 20 % greater than the birefringence of rutile and even outperforms the excitonic maximum anisotropy of CsPbBr$_3$ perovskite[18], as seen in Figure 4b. Therefore, Figure 4 demonstrates that As$_2$S$_3$ combines both high refractive index and giant optical anisotropy, which are the most crucial factors in the state-of-the-art nanophotonics.

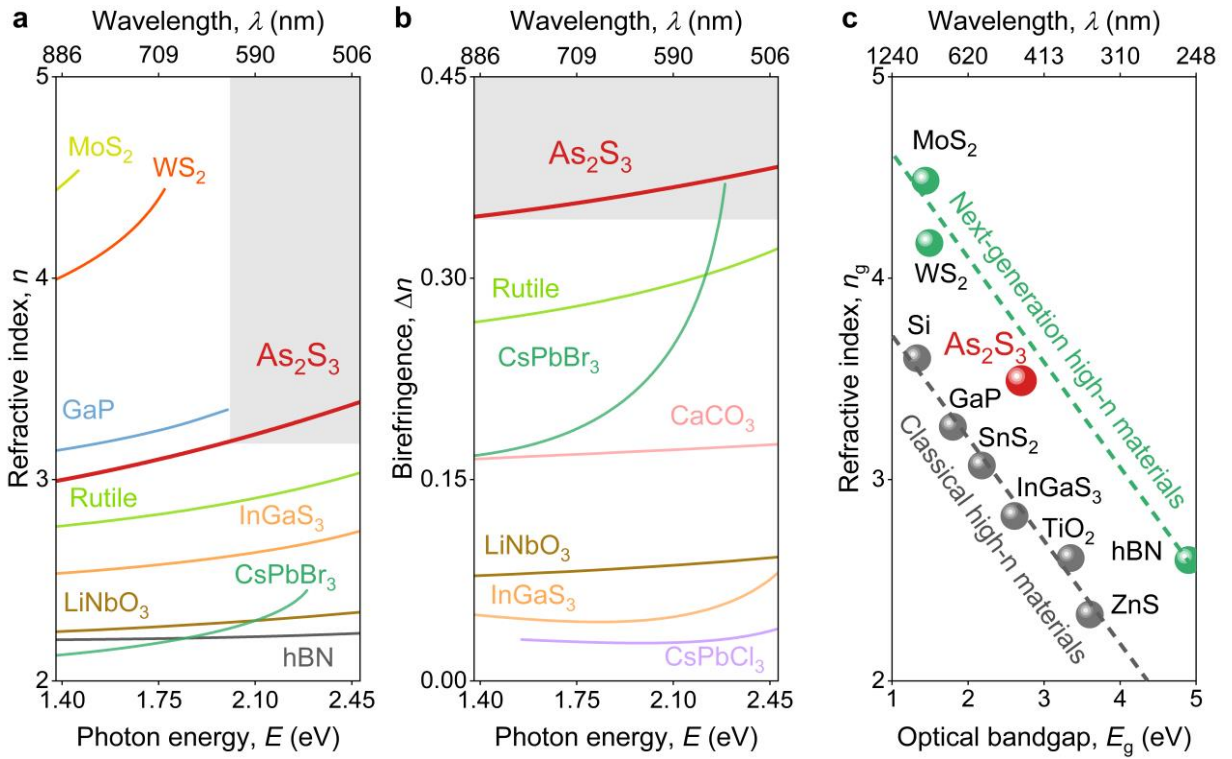

**Figure 4 | As$_2$S$_3$ in a family of high refractive index and birefringent materials. a,** Refractive index and **b,** the birefringence of van der Waals As$_2$S$_3$ and conventional photonic materials in their transparency windows. **c,** Comparison of the maximum in-plane refractive index of van der Waals As$_2$S$_3$ in the transparency window with the established highly refractive materials.

**Unconventional true zero-order quarter-waveplates based on van der Waals As$_2$S$_3$**

The exceptional optical properties of As$_2$S$_3$ (see Figure 4) not only make this crystal promising for next-generation nanophotonics but also change the operation principle of classical optical elements. In order to demonstrate this, we investigated the waveplate characteristics of As$_2$S$_3$ flake. Traditionally, the retardance $\delta$ between the fast- and slow-axes of an anisotropic waveplate is determined by the simple expression $2\pi\Delta n t/\lambda$, where $\lambda$ is the wavelength of light in vacuum, $\Delta n$ defines the material's birefringence, and $t$ denotes the waveplate's thickness. However, this formula only holds for small values of $\Delta n$ since it disregards the light scattering at the faces of an anisotropic material. Due to the significant difference in the refractive index components along the principal directions of the waveplate, the phase accumulation due to the repeated Fabry-Perot reflections makes the full retardance deviate from the simplified formula (see Figure 5a and Supplementary Notes 8-9). As a result, the giant optical anisotropy enables quarter-wave retardance at multiple wavelengths and at a thickness that is lower than predicted by the simplified expression. For instance, our As$_2$S$_3$ operates as a true zero-order quarter-waveplate at two wavelengths (512 and 559 nm), and at 559 nm its thickness is lower than expected from the simplified expression. In contrast, the simplified equation predicts only single-wavelength operation at 522 nm (see Figure 5b). Here, we utilized a micro-transmittance scheme (see Figure 1c) at 512 and 559 nm to check this concept. The resulting transmittance maps in Figures 5d-g confirm the predicted quarter-waveplate behavior for the selected wavelengths. Furthermore, Figures 5d-g demonstrate that the polarization angles responsible for a quarter-waveplate mode differ from 45° with respect to the principal optical axes, unlike classical quarter-waveplates with 45° orientation[31]. This effect also originates from the giant optical anisotropy, which results in unequal absolute values of transmission amplitudes, *A* and *B*, for the light polarized along principal directions (see Figure 5a). It explains our observations from Figure 5d-g. Finally, we would like to note that our quarter-waveplate has an extremely small thickness of 345 nm compared to the previous record-holder of ferrocene-based true zero-order quarter-waveplate with 1071 nm thickness operating at 636 nm wavelength[31]. Hence, our device has about threefold improvement in size, which brings us a step closer to miniaturized next-generation optical elements.

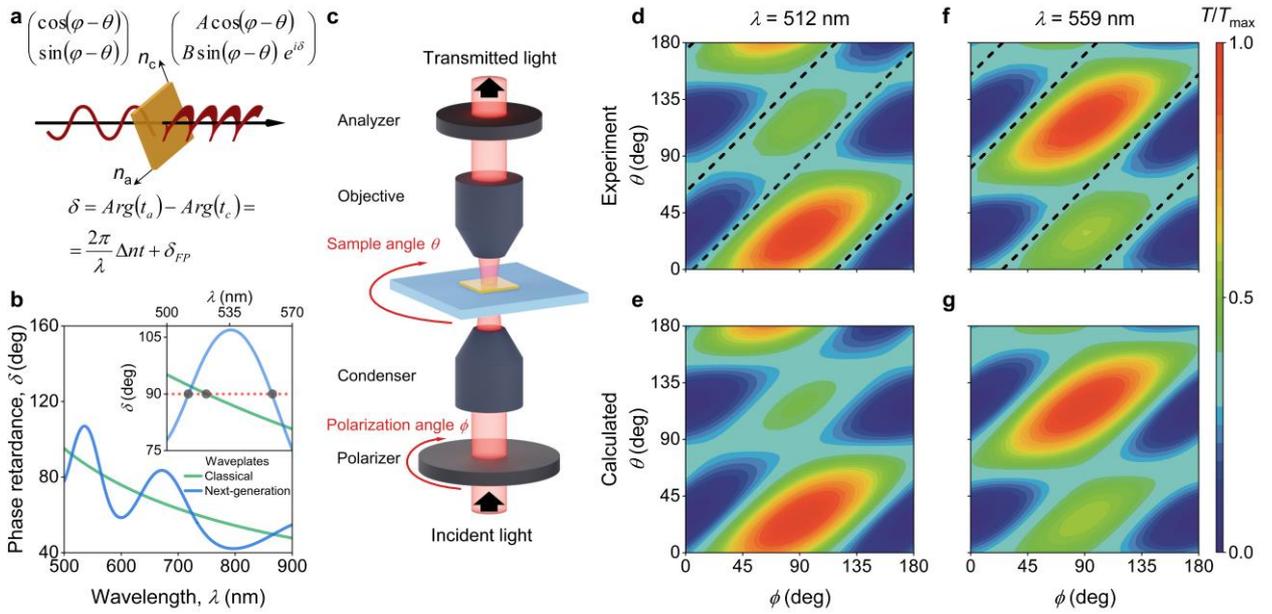

**Figure 5| True zero-order quarter-wave plates based on ultrathin van der Waals As$_2$S$_3$. a**, The concept of As$_2$S$_3$ waveplate: a combination of "classical" phase accumulation and Fresnel contribution arising from giant optical anisotropy. **b,** The comparison of phase retardance between classical and As$_2$S$_3$-based true-zero order waveplates. **c,** Schematic representation of the experimental setup. Measured polarized transmittance countourplot at **d,** 512 nm and **e,** 559 nm. The dashed lines show the quarter-waveplate operation regime. The data are normalized to the maximum values for each wavelength of transmitted light throughout the figure. Transmittance calculations are based on the anisotropic dielectric function presented in Figure 3e at **f,** 512 nm and **g,** 559 nm.

## Conclusion

In summary, we provide a new route for exploring anisotropic vdW materials by comparing their crystal structures and bandgaps. In combination with optical characterization, it becomes a convenient tool for a quick assessment of promising vdW materials for anisotropy-based applications[13–19]. Our approach reveals that As$_2$S$_3$ is a perfect vdW material for visible range nanophotonics with the largest in-plane optical anisotropy, high refractive index, and zero optical losses. These properties enrich photonic applications with a variety of novel possibilities at the nanoscale. For example, we designed an ultrathin two-wavelength As$_2$S$_3$-based quarter-wave plate, which is three times more compact than the thinnest single-wavelength quarter-waveplate[31]. Furthermore, our anisotropy analysis can be used beyond vdW materials. For instance, our assay predicts large anisotropy for non-vdW Sr$_{9/8}$TiS$_3$, which recently was discovered to have colossal optical anisotropy in the near-infrared range[17]. Besides, anisotropy of mechanical, electronic, optical, and other properties are closely connected, which allows using the proposed method for other topics. Indeed, As$_2$S$_3$ also demonstrates high mechanical anisotropy[48,49] in addition to the found giant optical anisotropy in our work. Therefore, our findings can lead to the rapid development of low-symmetry materials[34–37] by establishing a milestone for their anisotropy evaluation.

## Author Contributions

A.S.S. and G.A.E. contributed equally to this work. G.A.E., A.V.A., V.S.V., and K.S.N. suggested and directed the project. A.S.S., G.A.E., M.K.T., D.V.G., A.V.S., D.M.T., M.S.M., S.M.N., and D.A.G. performed the measurements and analyzed the data. A.N.T. and D.A.G. prepared the samples. O.G.M., A.M., K.V.K., I.K., and A.A.V. provided theoretical support. A.S.S. and G.A.E. wrote the original manuscript. A.S.S., G.A.E., D.A.G., A.A.V., A.V.A., V.S.V., and K.S.N. reviewed and edited the paper. All authors contributed to the discussions and commented on the paper.

## Competing Interests

The authors declare no competing interests.

## Methods

**Sample preparation.** Bulk synthetic $As_2S_3$ crystals were purchased from 2d semiconductors (Scottsdale) and exfoliated on top of Si, Si/SiO$_2$, quartz and Schott glass substrates at a room temperature by commercial scotch tapes from Nitto Denko Corporation (Osaka, Japan). Prior to exfoliation, the corresponding substrates were subsequently cleaned in acetone, isopropanol alcohol, and deionized water, and then, subjected to oxygen plasma (O$_2$) to remove the ambient adsorbates.

**Atomic-force microscopy characterization.** The thickness of $As_2S_3$ flakes was accurately characterized by an atomic force microscope (NT-MDT Ntegra II) operated in contact mode at ambient conditions. AFM measurements were acquired using silicon tips (ETALON, HA_NC ScanSens) with a head curvature radius of < 10 nm, a spring constant of 3.5 N/m and a resonant frequency of 140 kHz. Gwyddion software was used for image processing and quantitative analysis.

**X-ray diffraction analysis.** X-ray diffraction analysis of $As_2S_3$ single crystal was performed on a Bruker D8 QUEST diffractometer with a Photon III CMOS area detector using Mo Kα radiation ($\lambda$ = 0.71073 Å) focused by a multilayer Montel mirror. Full data set was collected at 293 K within $\varphi$- and $\omega$-scans applying sample-to-detector distance of 80 mm and 100 mm to improve the precision of refined unit cell parameters. Raw data were indexed with cell_now and integrated using SAINT from the SHELXTL PLUS package[52,53]. Absorption correction was performed using a numerical method based on crystal shape as implemented in SADABS. Crystal structure was solved by direct methods and refined anisotropically with the full-matrix F2 least-squares technique using SHELXTL PLUS. Further details of the data collection and refinement parameters are summarized in Supplementary Table 1. Selected interatomic distances and bond angles are listed in Supplementary Table 2. It is worth noting that the crystal structure of monoclinic $As_2S_3$ was previously reported[54,55] in non-conventional unit cell setting, which can be transformed to conventional setting by $\begin{pmatrix} 0 & 0 & 1 \\ 0 & -1 & 0 \\ 1 & 0 & 0 \end{pmatrix}$ matrix. Unit cell parameters and atomic positions reported in present work were determined with higher precision (Supplementary Table 3). CSD reference number 2258216 contains supplementary crystallographic data for this paper. These data can be obtained free of charge from the Cambridge Crystallographic Data Centre via www.ccdc.cam.ac.uk/data_request/cif.

**First-principle calculations.** Electronic bandstructure calculations were performed using the screened hybrid functional HSE06 with 25% of mixing as implemented in Vienna *ab initio* simulation package (VASP) code[56,57]. The core electrons are described with projector augmented wave (PAW) pseudopotentials treating the As 4s and 4p and the S 3s and 3p electrons as valence. A kinetic energy cutoff for the plane-wave basis was set to 350 eV. To calculate bandstructure we generated a path in reciprocal space using Spglib and SeeK-path and used a standardized primitive cell by conventions of Spglib. Optical properties of $As_2S_3$ were calculated using HSE06 hybrid functional. For this we used Γ-centered k-points mesh sampling the Brillouin zone with a resolution of 2π·0.05 Å$^{-1}$. Optical properties were calculated within GW approximation on wavefunctions calculated using HSE06 hybrid functional using the VASP code. For this, we obtained ground-state one-electron wavefunctions from HSE06 and used them to start the GW routines. Finally, we calculated the imaginary and real parts of the frequency-dependent dielectric function within GW approximation.

**Angle-resolved micro-transmittance.** The spectroscopic transmittance was measured in the 500–900 nm spectral range on an optical upright microscope (Zeiss Axio Lab.A1) equipped with a halogen light source, analyzer, polarizer, and grating spectrometer (Ocean Optics QE65000) coupled by optical fiber (Thorlabs M92L02) with core diameter 200 μm. The transmitted light was collected from a spot of <15 μm using an objective with ×50 magnification and numerical aperture N.A. = 0.8 (Objective "N-Achroplan" 50×/0.8 Pol M27). The detailed description of micro-transmittance setup can be found in publication[58].

**Imaging Mueller matrix ellipsometry.** A commercial Accurion nanofilm_ep4 ellipsometer (Accurion GmbH) was used to measure 11 elements of the Mueller matrix ($m_{12}$, $m_{13}$, $m_{14}$, $m_{21}$, $m_{22}$, $m_{23}$, $m_{24}$, $m_{31}$, $m_{32}$, $m_{33}$, $m_{34}$). The measurements were carried with a 5° sample rotation angle step, 550 nm incident light wavelength and 50° incident angle in rotation compensator mode.

**Scanning near-field optical microscopy.** Near-field imaging was performed using a commercially available scattering type Scanning Near Field Optical Microscope (neaSNOM), which allows to simultaneously scan the topography of the sample along with amplitude and phase of the near-field signal. For the illumination of the sample we used tunable Ti:Sapphire laser (Avesta) with a wavelength in the spectral range of 700–1000 nm. The measurements were conducted using reflection mode. As a scattering probe we used a platinum/iridium5 ($PtIr_5$) coated AFM tip (ARROW-NCPt-50, Nanoworld) with a resonant frequency of about 275 kHz and a tapping amplitude of 100 nm.

## Data Availability

The datasets generated during and/or analysed during the current study are available from the corresponding author upon reasonable request.